# ON THE NATURE OF RAPIDLY ROTATING SINGLE EVOLVED STARS

R. Rodrigues da Silva, B. L. Canto Martins, and J. R. De Medeiros
Departamento de Física Teórica e Experimental, Universidade Federal do Rio Grande do Norte, Campus Universitário, Natal RN, Brazil; renan@dfte.ufrn.br

## ABSTRACT

We present an analysis of the nature of the rapidly rotating, apparently single giant based on rotational and radial velocity measurements carried out by the CORAVEL spectrometers. From the analyzed sample, composed of 2010 spectroscopic, apparently single, evolved stars of luminosity classes IV, III, II, and Ib with spectral types G and K, we classified 30 stars that presented unusual, moderate to rapid rotation. This work reports, for the first time, the presence of these abnormal rotators among subgiant, bright giant, and Ib supergiant stars. To date, this class of stars was reported only among giant stars of luminosity class III. Most of these abnormal rotators present an *IRAS* infrared excess, which, in principle, can be related to dust around these stars.

## 1. INTRODUCTION

The rotational behavior of single subgiant, giant, bright giant, and Ib supergiant stars along the spectral range F, G, and K is now well established. Essentially, for F-type single evolved stars the rotational velocity ranges from a few km s$^{-1}$ to about one hundred times the rotation rate of the Sun, whereas G and K single evolved stars are essentially slow rotators, with rotation decreasing smoothly from early-G to late-K types (Gray 1989; De Medeiros et al. 1996; De Medeiros & Mayor 1999). Nevertheless, the literature reports a growing list of apparently single G- and K-type giant stars, namely stars of luminosity III, that violate the general rule with rotational velocities as high as 80 km s$^{-1}$ (Bidelman & MacConnell 1973; Fekel & Scarfe 1986; Balona 1987; Fekel & Balachandran 1993; De Medeiros & Mayor 1999; Fekel et al. 1987). To date, the nature of such an abnormal rotation is not yet well understood. For instance, different authors have suggested that these rapid rotators are coalesced W UMa binaries rather than single stars, following the evolutionary status proposed for FK Comae single stars (e.g., Bopp & Stencel 1981). Following a different approach, Simon & Drake (1989) proposed that the enhanced rotation of single giant stars could reflect an angular momentum sudden dredge-up from the stellar interior. Peterson (1983) suggested that planets could be the source of enhanced rotation in evolved stars. In addition, Siess & Livio (1999b) have shown that the accretion of planets and brown dwarfs by giant stars could produce a spin-up of the star as a result of angular momentum deposition from the accreted companion. The discovery of hot Jupiters, giant planets orbiting their host stars at inner solar system distances, seems to offer observational support for this approach once these hot Jupiters are close enough to eventually be swallowed up by their host stars (e.g., Livio & Soker 2002; Carney et al. 2003; Massarotti et al. 2008).

More recently, Carlberg et al. (2011) reported the discovery of 28 new rapidly rotating single giant stars with rotational velocity $v \sin i$ ranging from 10.0 to 86.4 km s$^{-1}$ from a sample of approximately 1300 K giants, representing 2.2% of their whole sample. In a more recent study (Carlberg et al. 2012), these authors have found consistent evidence that the engulfment of planets of a few Jupiter masses by the host star could be the root cause of the enhanced rotation of single giant stars, which is in agreement with previous theoretical predictions (e.g., Siess & Livio 1999b).

This work brings an additional analysis on the nature of rapidly rotating single evolved stars on the basis of an unprecedented survey of rotation and radial velocities for evolved stars carried out by De Medeiros & Mayor (1999) and De Medeiros et al. (2002, 2014). These authors offer a unique set of $v \sin i$ and radial velocity measurements for F-, G-, and K-type stars of luminosity classes IV, III, II, and Ib, respectively, subgiant, giant, bright giant, and Ib supergiant, within the Bright Star Catalogue (Hoffleit & Jaschek 1982; Hoffleit et al. 1983) and an additional sample of F-, G-, and K-type stars of luminosity classes II and Ib from Egret (1980).

## 2. STELLAR WORKING SAMPLE

The present stellar sample consists of G- and K-type stars of luminosity classes IV, III, II, and Ib, respectively, subgiant, giant, bright giant, and Ib supergiant stars listed in the Bright Star Catalogue; G- and K-type stars of luminosity classes II and Ib from Egret (1980) with projected rotational velocities $v \sin i$ given by De Medeiros & Mayor (1999), De Medeiros et al. (2002), and De Medeiros et al. (2014); and constant radial velocities presented on the basis of at least two observations carried out with the CORAVEL spectrometer (Baranne et al. 1979). Indeed, the behavior of the probability $P(X^2)$, a parameter indicator of CORAVEL radial velocity variability, was taken into account to consider the selected stars as single. As shown by Duquennoy et al. (1991), the parameter $P(X^2)$ should strongly peak toward very small values for binary stars and be flat from approximately 0.10 to 1 for single stars, namely those stars showing constant radial velocity. Figure 1 shows the distribution $P(X^2)$ for the stars classified as single or apparently single and as spectroscopic binaries by De Medeiros & Mayor (1999) and De Medeiros et al. (2002, 2014), which strictly follow the results of Duquennoy et al. (1991). The spectroscopic binaries show essentially $P(X^2)$ smaller than 0.10, whereas the single or apparently single stars show a flat behavior for $P(X^2)$ with values larger than 0.10 to 1. Let us also underline that, according to Duquennoy et al. (1991), for single stars the CORAVEL parameter E/T, which represents the ratio between the typical error for one single radial velocity measurement and the radial velocity dispersion (rms), is close to unity, whereas for binary stars, such a parameter spreads

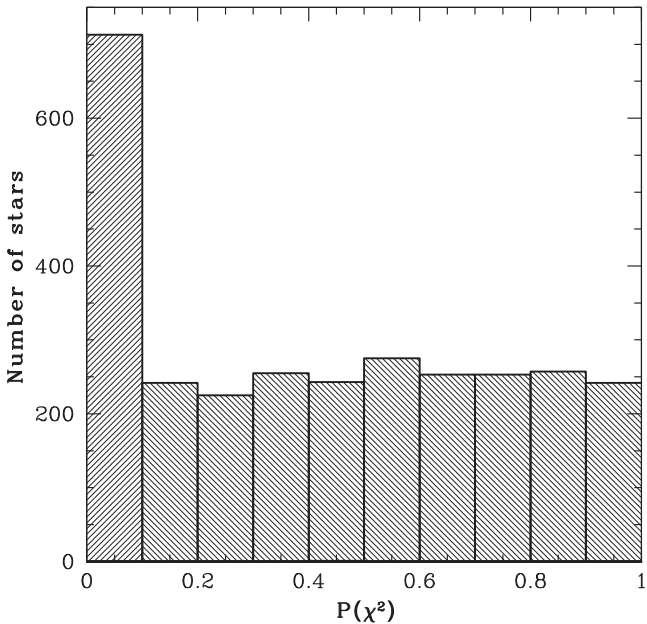

**Figure 1.** Distribution of the $P(X^2)$ probability for single and binary evolved stars from De Medeiros & Mayor (1999) and De Medeiros et al. (2002, 2014). The histogram contains 2010 G- and K-type, apparently single, evolved stars with slow rotation, the 30 rapid rotator single evolved stars referred in this study, and 713 spectroscopic binary systems. The distribution of $P(X^2)$ for constant radial velocity is rather flat from 0.1 to 1, while variable stars strongly peak toward values between 0 and 0.1.

toward values greater than 1 according to the amplitude range of the radial velocity of the binary.

For a complete discussion on the observational procedure, calibration, and error analysis, readers are referred to Duquennoy (1987), Duquennoy et al. (1991), De Medeiros & Mayor (1999), and De Medeiros et al. (2002, 2014). However, let us recall a few important points. The $v \sin i$ and radial velocity measurements are based on observations carried out with the two CORAVEL spectrometers (Baranne et al. 1979) mounted on the 1 m Swiss telescope at the Haute-Provence Observatory, Saint Michel (France), and the 1.54 m Danish telescope at the ESO, La Silla (Chile). Projected rotational velocity $v \sin i$ was obtained through an appropriate calibration of the widths of cc-dips, as described by De Medeiros & Mayor (1999), with a typical uncertainty of approximately 1.0 km s$^{-1}$ for subgiant and giant stars exhibiting $v \sin i$ lower than 30 km s$^{-1}$. For bright giants and Ib supergiants in particular, De Medeiros & Mayor (1990) developed a specific calibration for $v \sin i$, which takes into consideration the contribution of macroturbulence for the CORAVEL cross-correlation dip as a function of spectral type, based on the measurements of macroturbulence velocities estimated by Gray & Toner (1986) and Gray & Toner (1987). For the $v \sin i$ measurements of bright giants and Ib supergiants, De Medeiros & Mayor (1999) conservatively assumed an uncertainty of 2.0 km s$^{-1}$. For faster rotators, those with $v \sin i$ higher than 30 km s$^{-1}$, the estimations of De Medeiros & Mayor (1999) indicate an uncertainty of approximately 10% regardless of luminosity class.

Let us turn now to the definition of our working sample, namely G and K spectroscopic, apparently single, evolved rapid rotator stars, hereafter ERR stars. We have selected from De Medeiros & Mayor (1999) and De Medeiros et al. (2002, 2014) all the G and K subgiant, giant, bright giant, and Ib supergiant stars, namely G- and K-type stars of luminosity classes IV, III, II, and Ib, that present a constant radial velocity, as defined in the preceding paragraph, with a rotational velocity $v \sin i$ of 10 km s$^{-1}$ or larger. Essentially, the stars in the corresponding luminosity classes IV, III, II, and Ib are redward of $(B - V) = 0.55, 0.70, 0.65,$ and 0.70, respectively. These cutoffs in color correspond to clear regions where evolved stars show two very distinct rotational behaviors: single stars blueward of these colors present a large scatter in $v \sin i$ from a few km s$^{-1}$ to about one hundred times the solar rotation rate, whereas single stars redward of these colors are typically slow rotators (De Medeiros et al. 1996, 2002). The defined cutoff in $v \sin i$ is the same one applied by Carlberg et al. (2011) and Fekel (1997), which is slightly more conservative than that applied by Drake et al. (2002) and Massarotti et al. (2008), who also identified a few rapid rotator single giant stars. Our defined cutoff is at least four times larger than the observed mean rotational velocity $v \sin i$ for single evolved stars in the considered luminosity classes and spectral range given by De Medeiros et al. (1996). On the basis of these criteria, we identified a total of 30 ERR stars, as described in Table 1, corresponding to 17 stars of luminosity classes IV, III, II, and Ib from the Bright Star Catalogue and 13 G and K stars of luminosity classes II and Ib from Egret (1980). Three of the reported stars (HD 31993, HD 37434, and HD 112989) are classified in the literature as chromospherically active (Bidelman & MacConnell 1973; Fekel & Scarfe 1986; Balona 1987) but also as presenting spectroscopic, apparently single, behavior and moderate to rapid rotation. As one observes from Table 1, all the referred stars have a probability $P(X^2)$ larger than 0.10 and E/T values close to or lower than unity.

## 3. RESULTS

The first interesting finding that emerges from this work is the presence of 21 ERR stars among subgiant, bright giant, and Ib supergiant stars; only once to date has this class of rapid rotators been reported among only giant stars. Indeed, the whole sample is composed of 1 subgiant, 9 giants, 8 bright giants, and 12 Ib supergiants. As underlined in the previous section, we define as ERR stars those G- and K-type evolved stars of luminosity classes IV, III, II, and Ib redward of $(B-V) = 0.55, 0.70, 0.65,$ and 0.70, respectively, with a rotational velocity $v \sin i$ of 10 km s$^{-1}$ or larger. The distribution of these stars by luminosity class and $T_{\rm eff}$ is displayed in Figure 2, from which one observes that, at least in the luminosity classes III, II, and Ib, the ERR stars are located in the same interval of $T_{\rm eff}$. The locations of 17 of the previously mentioned rapid rotator stars in the HR diagram are shown in Figure 3, in which, in spite of the large error bars in the luminosity of some stars, one observes that their masses range from approximately 1.5 $M_\odot$ to at least 7 $M_\odot$. Indeed, 13 stars from the whole sample of 30 targets (BD +25 4819, BD +31 2471, BD +59 12, HD 232862, HD 70046, HD 99576, HD 95393, HD 84315, HD 87323, HD 101314, HD 149900, HD 4362, HD 192078) are not included in Figure 3 because they have no reliable parallax measurements listed in the literature. In Figure 3, evolutionary tracks at [Fe/H] = 0 are from Girardi et al. (2000) with luminosities derived from the HIPPARCO parallaxes (ESA 1997).

**Table 1**
Stellar Parameters, Including CORAVEL Projected Rotational Velocity $v \sin i$, for 30 Rapidly Rotating Single Evolved Stars

| ID | Spectral Type | $(B-V)$ | $T_{\rm eff}$ (K) | $v \sin i$ (km s$^{-1}$) | E/T | $P(\chi^2)$ | $N$ | $V - [12]$ | $V - [25]$ |
|---|---|---|---|---|---|---|---|---|---|
| Bright Star Catalogue | | | | | | | | | |
| HD 4362[b] | G0Ib | 1.09 | 4676 | 10.7 | 0.85 | 0.61 | 6 | 2.56 ± 0.07 | 2.68 ± 0.60 |
| HD 13994[b] | G7III | 1.05 | 4750 | 11.5 | 0.70 | 0.84 | 8 | 2.57 ± 0.05 | 2.61 ± 0.33 |
| HD 31910[b] | G0Ib | 0.92 | 5005 | 11.7 | 0.96 | 0.52 | 11 | 2.13 ± 0.01 | 2.18 ± 0.06 |
| HD 31993[a,b] | K2III | 1.28 | 4349 | 31.1 | 1.03 | 0.40 | 32 | 3.20 ± 0.14 | 3.17 ± 1.54 |
| HD 37434[a,b] | K2III | 1.16 | 4553 | 65.3 | 1.15 | 0.20 | 12 | 2.90 ± 0.05 | 2.94 ± 0.24 |
| HD 66011 | G0IV | 0.57 | 6002 | 13.6 | 0.69 | 0.75 | 5 | 1.38 ± 0.26 | 2.39 ± 0.00 |
| HD 66812 | G8II | 1.01 | 4825 | 10.6 | 0.70 | 0.49 | 2 | 2.37 ± 0.11 | 1.96 ± 1.30 |
| HD 74006 | G7Ib-II | 0.94 | 4963 | 11.8 | 1.21 | 0.23 | 2 | 2.15 ± 0.01 | 2.14 ± 0.07 |
| HD 101570[b] | G3Ib | 1.15 | 4570 | 21.4 | 1.39 | 0.16 | 2 | 2.68 ± 0.02 | 2.69 ± 0.13 |
| HD 112989[a,b] | G9IIICH-2F | 1.17 | 4535 | 11.0 | 0.44 | 0.66 | 2 | 2.80 ± 0.01 | 2.81 ± 0.11 |
| HD 121107[b] | G5III | 0.84 | 5183 | 14.5 | 0.61 | 0.95 | 10 | 1.97 ± 0.08 | 2.29 ± 0.00 |
| HD 137465[b] | G2II | 1.09 | 4676 | 10.9 | 0.99 | 0.41 | 5 | 2.73 ± 0.06 | 2.82 ± 0.33 |
| HD 170845[b] | G8III | 1.01 | 4825 | 11.8 | 1.29 | 0.20 | 2 | 2.33 ± 0.02 | 2.41 ± 0.15 |
| HD 176884[b] | G6III | 1.29 | 4333 | 13.4 | 1.31 | 0.20 | 6 | 3.07 ± 0.02 | 3.16 ± 0.17 |
| HD 178937[b] | G2III | 1.02 | 4806 | 30.7 | 1.41 | 0.16 | 2 | 2.47 ± 0.13 | 2.95 ± 0.00 |
| HD 202314[b] | G2Ib | 1.09 | 4676 | 14.6 | 1.07 | 0.33 | 9 | 2.60 ± 0.07 | 2.65 ± 0.50 |
| HD 223460[b] | G1IIIe | 0.79 | 5307 | 21.5 | 0.86 | 0.86 | 40 | 0.57 ± 0.57 | 1.92 ± 0.55 |
| Egret (1980) | | | | | | | | | |
| HD 70046 | G3Ib | 0.92 | 5005 | 25.4 | 0.49 | 0.63 | 2 | ... | ... |
| HD 84315[b] | G6Ib-II | 1.00 | 4844 | 10.3 | 0.88 | 0.39 | 2 | 3.25 ± 0.33 | 4.39 ± 0.00 |
| HD 87323[b] | G2Ib | 1.10 | 4659 | 10.4 | 0.36 | 0.72 | 2 | 2.50 ± 0.26 | 3.52 ± 0.00 |
| HD 95393[b] | G3Ib | 1.44 | 4087 | 10.6 | 1.47 | 0.14 | 2 | 3.38 ± 0.10 | 3.53 ± 0.98 |
| HD 99576 | G3/5Ib | 1.32 | 4284 | 10.4 | 0.24 | 0.81 | 2 | ... | ... |
| HD 101314[b] | G2Ib | 1.08 | 4695 | 10.2 | 1.12 | 0.26 | 2 | 2.19 ± 0.30 | 3.69 ± 0.00 |
| HD 149900 | G0/2Ib | 0.76 | 5387 | 10 | 0.71 | 0.48 | 2 | ... | ... |
| HD 192078[b] | G5II | 1.53 | 3926 | 11.4 | 1.04 | 0.36 | 18 | 3.62 ± 0.10 | 5.32 ± 0.00 |
| HD 206121[b] | G5II | 0.83 | 5207 | 14.9 | 0.85 | 0.54 | 4 | 2.01 ± 0.27 | 3.09 ± 0.00 |
| HD 232862 | G8II | 0.87 | 5114 | 20.6 | 0.91 | 0.59 | 8 | ... | ... |
| BD+25 4819 | K0II | 1.08 | 4695 | 14.9 | 0.98 | 0.46 | 6 | ... | ... |
| BD+31 2471 | K0II | 0.92 | 5005 | 13.8 | 0.78 | 0.55 | 3 | ... | ... |
| BD+59 12 | G0II | 1.02 | 4806 | 20.9 | 0.00 | 1.00 | 2 | ... | ... |

[a] Chromospherically active stars with CORAVEL constant radial velocity.
[b] IR excess.

### 3.1. On The Nature of the Rapid Rotator Single Evolved Stars

As already underlined in this work, to date, the nature of the ERR stars is essentially unknown despite the number of mechanisms proposed by different authors, which includes a possible angular momentum sudden dredge-up from the stellar interior (Simon & Drake 1989), binary coalescence (e.g., Bopp & Stencel 1981), and the accretion of substellar companions (Siess & Livio 1999a, 1999b; Peterson 1983). Nevertheless, these mechanisms are applied only to solar-mass stars (Siess & Livio 1999a, 1999b) or W UMa binaries, which have a total system mass between 1 and 3 solar masses (Jiang et al. 2010).

As a first step in this study to understand the nature of ERR stars, we analyzed the behavior of the rotational velocity $v \sin i$ as a function of effective temperature for the rapid rotators in comparison to spectroscopic binaries with G- and K-type evolved components of luminosity classes IV, III, II, and Ib, listed by De Medeiros & Mayor (1999) and De Medeiros et al. (2002, 2014), which also present a $v \sin i$ of 10 km s$^{-1}$ or larger, namely, rapidly rotating binary systems according to our criterion. Figure 4 displays the $v \sin i$ values for our sample of 30 ERR stars and for the G- and K-type spectroscopic binaries. Interestingly, the $v \sin i$ versus $T_{\rm eff}$ distribution for both families of stars appears to have similar behavior. Because of this apparent common characteristic in the distribution of $v \sin i$ for the G and K rapidly rotating single evolved stars that have been compared with rapidly rotating binary systems of similar spectral types, we have performed a Kolmogorov–Smirnov (KS) test (Press et al. 1992), which calculates the probability that two data sets are derived from the same parent distribution, to verify if the present $v \sin i$ data sets for our 30 ERR stars and the rapid rotator binary stars are significantly different from one another. Figure 5 shows the cumulative distributions for the $v \sin i$ of both families of stars, for the luminosity classes III, II, and Ib. According to the KS test, the distributions are dissimilar if a probability of zero is obtained, whereas the unit probability indicates that they are similar. The non-zero probability values of 0.451, 0.577, and 0.016 obtained for luminosity classes III, II, and Ib, respectively, are consistent with the distributions of $v \sin i$ of the 30 ERR stars and rapidly rotating binaries being drawn from the same parent distribution. In this sense, as shown observationally by different authors (e.g., De Medeiros et al. 2002), the process of synchronization between rotation and orbital motions due to tidal interaction in binary systems with G- and K-type evolved components increases the rotation of the binary systems up to approximately 15 times the mean rotational velocity of single giants.

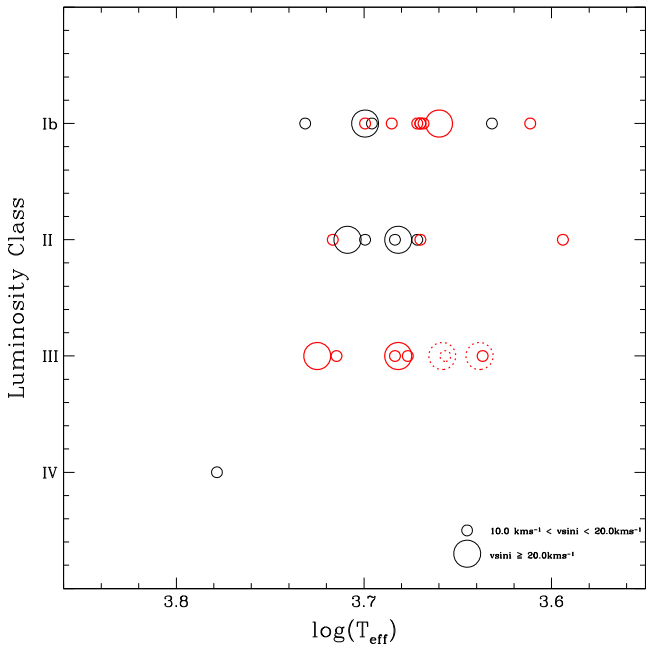

**Figure 2.** Distribution of the 30 single evolved stars with rapid rotation referred to in this study, by luminosity class and $T_{\rm eff}$. Red and black circles refer to stars with and without *IRAS* infrared excess, respectively. Chromospherically active stars are indicated by dashed circles.

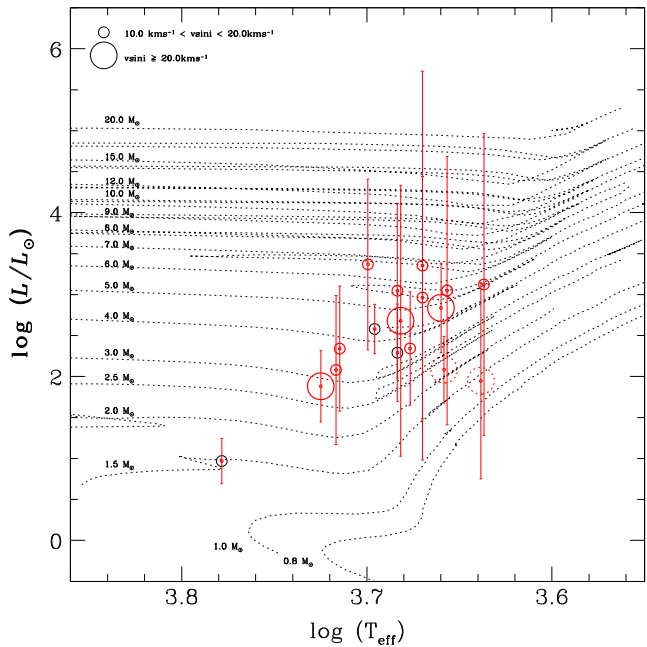

**Figure 3.** Distribution of rapidly rotating single evolved stars in the HR diagram. In this figure only 17 stars from the entire sample with reliable parallax measurements are reported. Luminosities have been derived from the HIPPARCO parallaxes (ESA 1997). Evolutionary tracks at [Fe/H] = 0 are shown for stellar masses between 0.8 and 20.0 $M_\odot$ (Girardi et al. 2000). Red and black circles refer to stars with and without *IRAS* infrared excess. Chromospherically active stars are indicated by dashed circles.

Nevertheless, in the same physical basis, the torques that act on both stars and planets in a planetary system as a result of tidal interactions can also spin up the host star (e.g., Brown et al. 2011). For instance, these authors have found evidence that the stars WASP-18 and WASP-19 have been spun up

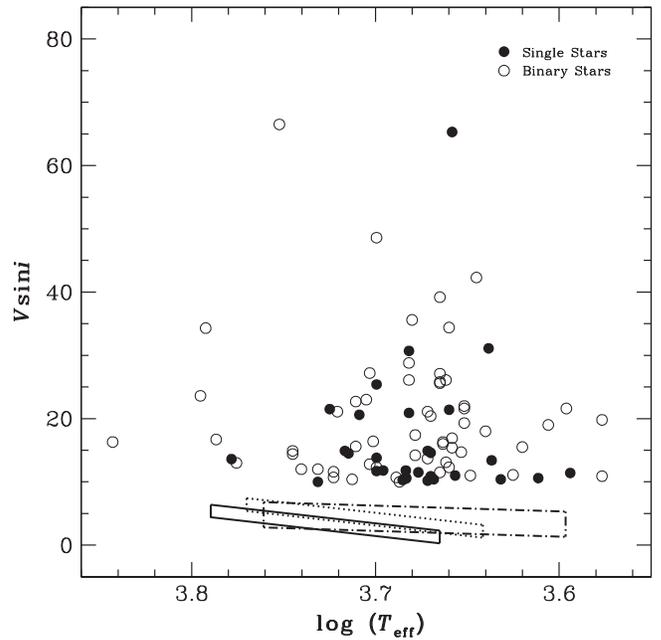

**Figure 4.** Projected rotational velocity $v \sin i$ vs. $T_{\rm eff}$ for the rapidly rotating single evolved stars (filled circles) and for rapidly rotating binary systems (open circles) with G- and K-type evolved components. The zones illustrated in the lower part of the figure represent the mean rotational velocity for evolved stars with luminosity classes IV (solid line), III (dashed line), and II–Ib (dashed–dotted line) given by De Medeiros et al. (1996).

through tidal interactions as their planets spiral toward their Roche limits.

We also analyzed the *IRAS* far-infrared (IR) behavior of the present sample of 30 ERR stars. Following the recipe by Walker & Cohen (1988), we have computed the *IRAS* colors for these stars in order to identify a possible IR excess in relation to the normal G and K giant, bright giant, and Ib supergiant stars. Figure 6 presents the computed *IRAS* colors $V–[12]$ and $V–[25]$ for 23 stars and, for comparative purposes, the best fit for the distribution of the *IRAS* colors for normal G and K evolved stars given by Cohen et al. (1987). A trend of an excess in far-IR emission is clearly observed for most of the 23 stars, a fact that may reflect the presence of warm dust close to these stars. The presence of debris disks at these evolutionary stages was, to date, only occasionally identified (e.g., Hillenbrand et al. 2008), despite the fact that, as shown by Bonsor et al. (2014), large quantities of dusty material can survive the stellar lifetime in the main sequence and be detected along the subgiant branch. On the other hand, the presence of dust near class III giant stars was reported by Zuckerman et al. (1995) and also in RS CVn tidally interacting binary systems (Busso et al. 1987; Busso et al. 1988; Scaltriti et al. 1993; Matranga et al. 2010). Scaltriti et al. (1993) proposed that the dust around RS CVn stars results from the accumulation of stellar winds driven by magnetic activity, whereas Matranga et al. (2010) follows the suggestion of Rhee et al. (2008) that the dust could originate from the collisions of planetary companions around those systems. For intermediate-mass giant and supergiant stars, in particular in the asymptotic giant branch, circumstellar dust was reported by different authors (e.g., Stencel et al. 1989; Young et al. 1993; Waters et al. 1994; Izumiura et al. 1996; Hashimoto & Izumiura 1998), but the origin of this material is not yet well established.

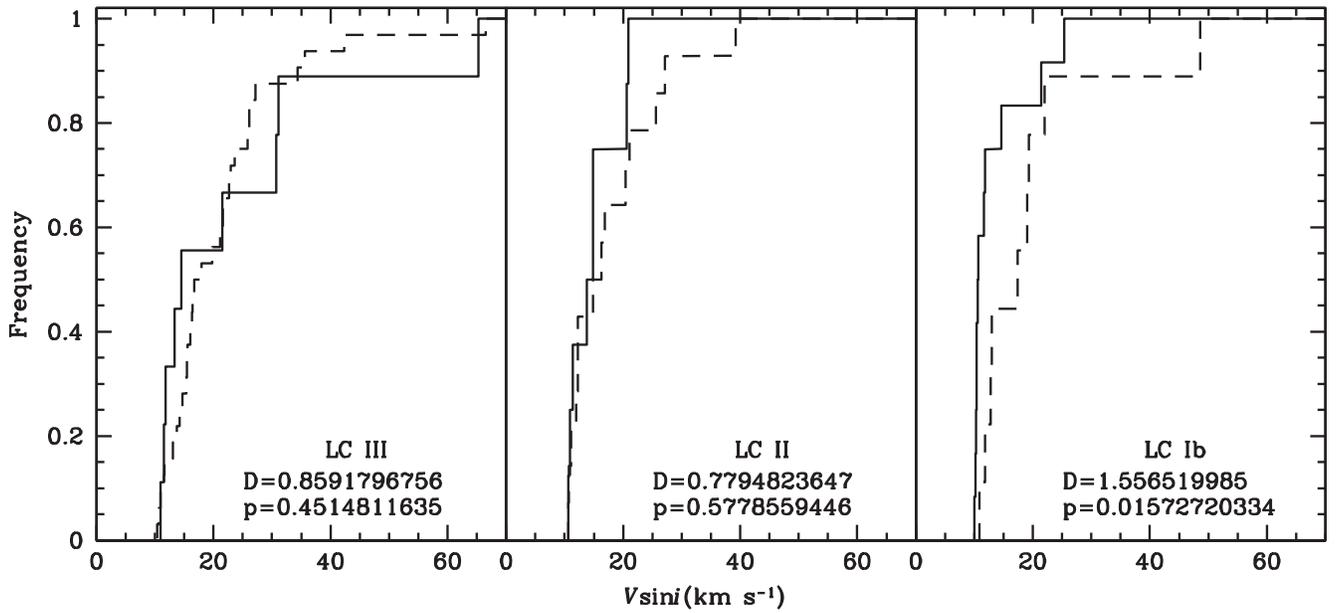

**Figure 5.** Cumulative distribution of the projected rotational velocity $v \sin i$ for the rapidly rotating single evolved stars and for rapidly rotating binary systems with G- and K-type evolved components. Solid and dashed curves represent single and binary stars, respectively. The distributions are separated by luminosity classes: III (left panel), II (middle panel), and Ib (right panel).

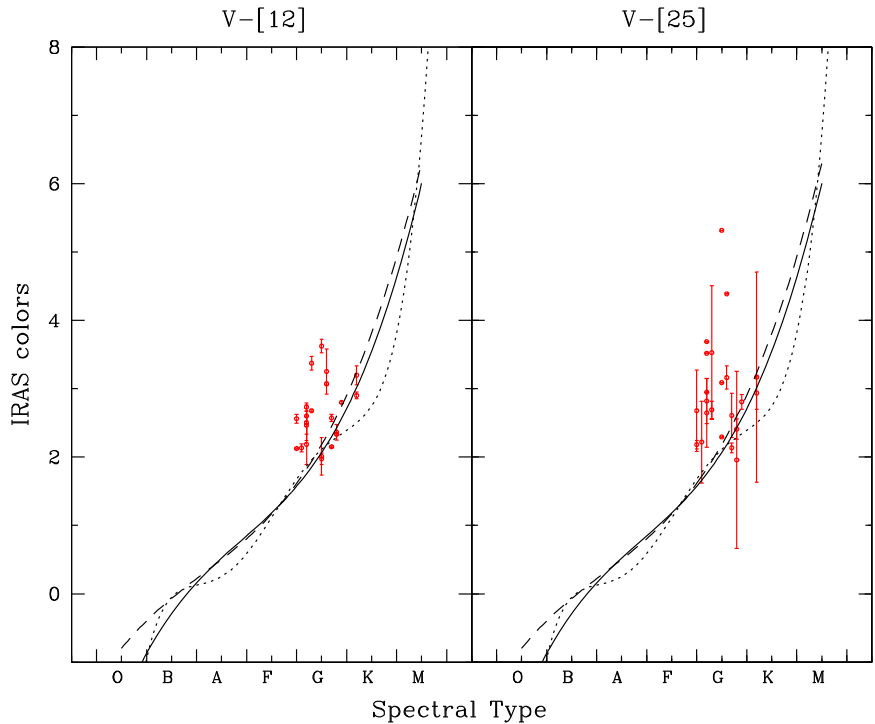

**Figure 6.** *IRAS* colors $V-[12]$ and $V-[25]$ for 23 G and K rapidly rotating single evolved stars (red symbols) of this work and, for comparative purposes, the best fit for the distribution of *IRAS* colors for G and K evolved stars with normal *IRAS* emission, given by Cohen et al. (1987), with dashed, solid, and dotted lines corresponding to the fit for classes Ib, II, and III, respectively.

## 4. CONCLUSIONS

The discovery of single G- and K-type giant stars with abnormally high rotation, typically giant stars that rotate excessively fast in relation to theoretical predictions, represents a real puzzle for stellar astrophysics today. In this study, we report the presence of these abnormal stars in classes of evolved stars other than the giants of luminosity class III, including the discovery of one subgiant, two bright giants, and five Ib supergiants listed in the Bright Star Catalogue. Adding these stars to the nine giants of luminosity class III from the catalog that were confirmed or discovered in this study, the total number of G- and K-type spectroscopic, apparently single, evolved stars in the Bright Star Catalogue with unusually rapid rotation amounts to 17 stars. This represents 0.8% of all G- and K-type apparently single evolved stars of luminosity classes IV, III, II and Ib listed in the Bright Star Catalogue. We also report the discovery of 13 G and K apparently single evolved stars of luminosity classes II and Ib listed by Egret ([1980](#)) that present abnormal enhanced rotation.

A trend of excess *IRAS* far-IR emission is clearly observed for most of the rapid rotator single evolved stars reported in this study. Such a fact offers an important additional piece of the puzzle on the nature of this abnormal rotation, irrespective of the luminosity class. Indeed, the mechanism producing such abnormal rotation, in principle, should be able to also yield warm dust around these stars. For low-mass stars, the mechanism of accretion of brown dwarfs and planets by giant stars, proposed by Siess & Livio ([1999a](#), [1999b](#)), may explain both the observed IR excess and the abnormal rotation. Actually, the reported incidence of debris disks surrounding evolved stars (e.g., Zuckerman et al. [1995](#); Melis et al. [2009](#)), in which the tidal ingestion of close-in planets by the host stars produces angular momentum deposition in addition to circumstellar dust, appears to support such an approach. Of course, the coalescence of contact binary systems during the initial ascent of the giant branch of the primary component is predicted to also produce angular momentum deposition in the coalesced system, which would resemble a late giant star with rapid rotation (e.g., Webbink [1976](#)), in addition to circumstellar dust. On the other hand, the stellar spin-up of stars as a result of tidal interactions with giant planets should also be considered (Brown et al. [2011](#)), but in such a scenario an IR excess is not necessarily expected. In this context, the rapid rotator single evolved star phenomenon may represent the final evolutionary stage of a coalescence scenario between a star and a low-mass stellar or substellar companion or the result of tidal interaction in planetary systems with close-in giant planets.

Although we are unable to point out the most plausible root cause for the abnormal enhanced rotation of single G- and K-type evolved stars, this study shows clear evidence that the presence of far-IR excess should be considered as an additional property for most of the targets reported in this stellar family. Additional experiments to further test this finding should consider, for example, an IR analysis in other IR bands to determine the nature of the material around these stars. In addition, it is mandatory to search for changes in specific abundance ratios, including the search for relative enhancement of refractory elements over volatiles, an aspect that could explain the difference between binary coalescence, planet accretion, or tidal interaction of star–star or star–substellar companions.

Research activities of the Stellar Board of the Federal University of Rio Grande do Norte are supported by CNPq and FAPERN Brazilian agencies. R.R.S. acknowledges a CNPq graduate fellowship and a CAPES/PNPD Post-Doctorate fellowship. We also acknowledge the INCT INEspaço for partial financial support. More information on the AASTeX macros package is available at http://aas.org/publications/aastex. For technical support, please write to aastex-help@aas.org.